\documentclass[pra,superscriptaddress,showpacs]{revtex4}
\usepackage{epsfig}
\usepackage{times}
\usepackage{amsmath}
\usepackage{amsfonts}
\usepackage{amssymb}
\usepackage[usenames]{color}
\usepackage{graphicx}
\newcommand{\beq}{\begin{equation}}
\newcommand{\eeq}{\end{equation}}
\newcommand{\bea}{\begin{eqnarray}}
\newcommand{\eea}{\end{eqnarray}}
\newcommand{\bec}{\begin{center}}
\newcommand{\enc}{\end{center}}
\newcommand{\bfr}{\begin{flushright}}
\newcommand{\efr}{\end{flushright}}
\newcommand{\la}{\langle}
\newcommand{\ra}{\rangle}
\newcommand{\dg}{\dagger}
\newcommand{\alp}{\alpha}

\newcommand{\om}{\omega}
\newcommand{\kap}{\kappa}
\newcommand{\gam}{\gamma}
\newcommand{\s}{\sigma}

\newcommand{\lam}{\lambda}

\newcommand{\ta}{\widetilde{a}}
\newcommand{\tb}{\widetilde{b}}
\newcommand{\td}{\widetilde{d}}
\newcommand{\te}{\widetilde{e}}

\newcommand{\tom}{\widetilde{\omega}}

\newcommand{\cF}{{\cal F}}
\newcommand{\cH}{{\cal H}}
\newcommand{\cN}{{\cal N}}

\newcommand{\cL}{{\cal L}} 
\newcommand{\cP}{{\cal P}} 
 
\begin{document}
\title{Entangling homogeneously broadened matter qubits 
in the weak-coupling cavity-QED regime}
\author{Kazuki Koshino}
  \affiliation{
   College of Liberal Arts and Sciences, Tokyo Medical and Dental
   University, Ichikawa, Chiba 272-0827, Japan
  }
 \author{Yuichiro Matsuzaki}
   \affiliation{
   NTT Basic Research Laboratories, NTT Corporation, 
   Kanagawa 243-0198, Japan
   }
\date{\today}
\begin{abstract}
In distributed quantum information processing, 
flying photons entangle matter qubits confined in cavities.
However, when a matter qubit is homogeneously broadened, 
the strong-coupling regime of cavity QED is typically required,
which is hard to realize in actual experimental setups.
Here, we show that a high-fidelity entanglement operation 
is possible even in the weak-coupling regime in which dampings
(dephasing, spontaneous emission, and cavity leakage) 
overwhelm the coherent coupling between a qubit and the cavity.
Our proposal enables distributed quantum information processing
to be performed using much less demanding technology than previously.
\end{abstract}
\maketitle

Distributed architecture is a promising approach for realizing scalable
quantum computation \cite{CEHM01a, Barrett:2005p363, Lim:2005p364,
BPH01a, Bose:1999p326, Feng:2003p334}.
Elementary nodes composed of a few qubits are networked to achieve scalable quantum computation.
The node separation 
can potentially suppress decoherence
induced by uncontrollable interactions between qubits. Moreover,
since the nodes are spatially separated, 
individual qubits can be easily addressed by the optical field.

A critical operation for realizing distributed quantum
computation is the entanglement operation 
(EO)~\cite{CEHM01a, Barrett:2005p363, Lim:2005p364,
BPH01a, Bose:1999p326, Feng:2003p334, laddetal2006hybrid, van2006hybrid,
azuma2009optimal}.
To construct an entire network,
qubits in distant nodes have to be coupled by EOs. 
Most EOs are based on photon interference, 
and the successful execution of an EO can be heralded
by detecting a photon at the target port.
This approach has been experimentally demonstrated 
using an ion trap system \cite{Moehring:2007p337}. 
If the EO fails, the two qubits involved should be initialized, 
which risks destroying the entanglement 
of other qubits generated by previous EOs.
Although EOs typically have such probabilistic properties, 
previous studies have revealed that 
only polynomial steps are required 
to construct large entangled states
\cite{Barrett:2005p363,Nielsen:2004p371,Duan:2005p369, Gross:2006p40,
YSJ01a}, such as a cluster state \cite{Raussendorf:2001p368}.
Moreover, by introducing a quantum memory to each node, 
EOs can be repeatedly performed until they are successful 
without destroying prior entanglement \cite{Benjamin:2006p358}.

EOs involve optical excitations of matter qubits.
However, the excited states are inherently noisy and
significantly degrade the target entanglement.
For example, nitrogen vacancy (NV) centers in diamond have 
promising properties such as a long coherence time 
at room temperature and optical addressability. 
Entanglement between an NV center and an emitted photon has been demonstrated
at a low temperature of about 7 K \cite{togan2010quantum}.
However, at room temperature,
this otherwise attractive system suffers from strong 
environmental dephasing originating from interactions with phonons
when the system is optically excited.
Consequently, it acquires a large homogeneous broadening 
of the order of THz \cite{fu2009observation}.
Therefore, in such an approach, NV centers can be used 
for distributed quantum computation only at low temperatures.

One way to overcome homogeneous broadening is to employ high-Q cavities.
Previous theoretical proposals of EOs
require strong coupling between a matter qubit and the cavity
when the matter qubit has large homogeneous broadening
\cite{CTSL01a, matsuzaki2011entangling,matsuzaki2011entanglementfinland}.
However, despite rapid advances in cavity fabrication technology, 
it is still very difficult to experimentally generate strong coupling between a matter qubit and a high-Q cavity.
To realize distributed quantum computation, 
it is thus essential to examine the possibility of performing an EO 
in the weak-coupling regime of cavity QED, 
where damping parameters such as the pure dephasing rate, 
the spontaneous emission rate, and the cavity decay rate
overwhelm the coherent coupling between the cavity and the qubit.

Here, we report that
high-fidelity entanglement can be generated between homogeneously broadened matter qubits
even in the weak-coupling regime of cavity QED.
Remarkably, both spontaneous emission of the qubit
and detuning between the photon and the qubit suppress 
environmental noise even for low-Q cavities,
and enable distributed quantum computation to be performed
using much less demanding technology than previously.
Moreover, since appropriate use of detuning has the potential 
to overcome the huge homogeneous broadening 
that is the main obstacle in using NV centers at high temperatures,
our analysis provides the possibility of using NV centers for
EOs at much higher temperatures than those of current 
experiments~\cite{togan2010quantum, sipahigil2011quantum}.

\begin{figure}[h]
\includegraphics[width=79mm]{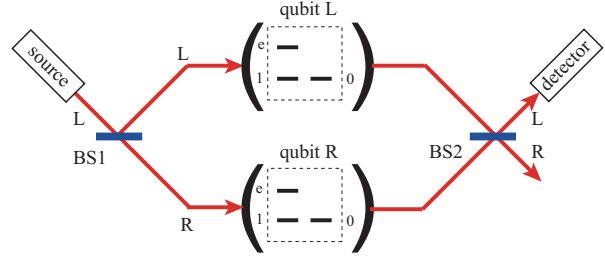}
\caption{Schematic view of the optical circuit. 
A single photon is split by a beam splitter (BS1)
and is sent to cavities that confine matter qubits with an L-type structure.
After interacting with the matter qubits, the photon 
is combined by another beam splitter (BS2).
When the photon reaches the target port and the detector clicks,
entanglement is generated between the remote matter qubits. 
}
\label{fig:MZ}
\end{figure}

An outline of the proposed scheme is as follows.
The matter qubit is the two ground states ($|0\ra$ and $|1\ra$)
of an L-type three-level system confined in a two-sided cavity.
$|0\ra$ is optically inactive, whereas $|1\ra$ is 
radiatively coupled to an excited state $|e\ra$
that is subject to level fluctuations due to environmental noise.
Two such qubits in cavities are placed symmetrically 
in a Mach--Zehnder interferometer (Fig.~\ref{fig:MZ}).
Initially, both qubits are prepared in $(|0\ra+|1\ra)/\sqrt{2}$
and a single photon tuned to the cavity frequency
is input from the left port of the first beam splitter (BS1).
The state vector of the system is given by
\beq
a_L^{\dag}(|00\ra+|01\ra+|10\ra+|11\ra)/2,
\label{eq:init}
\eeq
where $|mn\ra=|m\ra_L|n\ra_R$ denotes the two-qubit state vector
and $a_L^{\dag}$ ($a_R^{\dag}$) creates a
photon in the left (right) path.
The beam splitters divide a photon as 
$a_L^{\dag} \to (a_R^{\dag}+ia_L^{\dag})/\sqrt{2}$ and
$a_R^{\dag} \to (a_L^{\dag}+ia_R^{\dag})/\sqrt{2}$. 
For the interaction between the photon and the qubit,
when the qubit is in $|0\ra$ (empty cavity),
the input photon is perfectly transmitted through the cavity
due to resonance tunneling.
In contrast, when the qubit is in $|1\ra$, 
the matter qubit modifies the transmitted photon.
For example, as we show later,
the matter qubit may completely prevent transmission of the photon 
under some conditions.
Then, the photon--qubit interaction removes the terms
$a_L^{\dag}|10\ra$, $a_L^{\dag}|11\ra$,
$a_R^{\dag}|01\ra$ and $a_R^{\dag}|11\ra$.
In other words, the qubit state $|1\ra$ acts as a ^^ ^^ bomb''
in the interaction-free measurement~\cite{elitzur1993quantum}
in this case.
After the photon passes through the second beam splitter (BS2), 
the state vector is given by
 \beq
 -\frac{1}{\sqrt{8}}a_L^{\dag}|\phi_t\ra+
 \frac{i}{2}a_R^{\dag}|00\ra+
 \frac{i}{\sqrt{8}}a_R^{\dag}|\phi_1\ra,
 \eeq
where $|\phi_t\ra=(|01\ra-|10\ra)/\sqrt{2}$
and $|\phi_1\ra=(|01\ra+|10\ra)/\sqrt{2}$.
A photodetector is set to count the photons that exit from the left port of BS2.
The detector clicks with a maximum success probability of $1/8$, 
and the two qubits are then projected onto the target entangled state, $|\phi_t\ra$.

\begin{figure}[h]
\includegraphics[width=60mm]{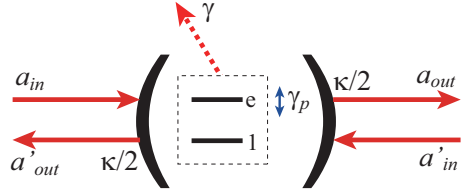}
\caption{Schematic view of the cavity QED system that we adopted. 
It consists of a matter qubit and a cavity. 
The incoming (outgoing) photon fields are denoted by
$\hat{a}_{in}$ and $\hat{a}'_{in}$ 
($\hat{a}_{out}$ and $\hat{a}'_{out}$).
Three types of damping are considered:
environmental pure dephasing ($\gamma_p$),
spontaneous emission ($\gamma$), 
and cavity photon leakage ($\kappa$).
}
\label{fig:cQED}
\end{figure}

We investigate the interaction 
between the photon and the qubit in a more quantitative manner.
Although the master equation has been used in previous 
analyses~\cite{matsuzaki2011entangling, matsuzaki2011entanglementfinland},
it is valid in principle only when the damping parameters
can be regarded as perturbations~\cite{arxivhornberger2009introduction}.
In contrast, here, we solve the Heisenberg equations of the overall system 
including the environment in a non-perturbative manner.
Consequently, our results are applicable to 
highly dissipative cases that include the weak-coupling regime.
We investigate a cavity QED system
in which a two-level matter qubit ($|1\ra$, $|e\ra$)
is confined in a two-sided cavity (Fig.~\ref{fig:cQED}).
The photon dynamics for the qubit state $|0\ra$ 
is obtained by removing the matter qubit.
This system is characterized by the following parameters:
the cavity frequency $\om_c$,
the qubit transition frequency $\om_q$,
the coherent coupling between the cavity and the qubit $g$,
the cavity decay rate $\kappa$,
the spontaneous emission rate of the qubit to non-cavity modes $\gamma$,
and the pure dephasing rate of the qubit $\gamma_p$.
The complex frequencies of the cavity and the qubit are defined by
$\tom_c=\om_c-i\kap/2$ and $\tom_q=\om_q-i(\gamma/2+\gamma_p)$.
We denote the destruction operators for the cavity photon and qubit
by $c$ and $\sigma(=|1\ra\la e|)$, respectively.
Their Heisenberg equations are given by
\bea
\frac{dc}{dt} &=& -i\tom_c
c-ig\s-i\sqrt{\kap/2}[a_{in}(t)+a'_{in}(t)],
\\
\frac{d\s}{dt} &=& -i\tom_q\s-igc-i\sqrt{\gamma}d_{in}(t)\nonumber \\
&&-i\sqrt{2\gamma_p}[e^{\dagger}_{in}(t)\s+\s e_{in}(t)],
\eea
where $d_{in}$ and $e_{in}$ denote the noise operators
respectively associated with spontaneous emission and pure dephasing, 
and $a_{in}$ and $a'_{in}$ are the incoming photon fields 
toward the cavity (see Fig.~\ref{fig:cQED}). 
The outgoing field operators are given by
\bea
a_{out}(t) &=& -a'_{in}(t)+i\sqrt{\kap/2}c(t),
\\
a'_{out}(t) &=& -a_{in}(t)+i\sqrt{\kap/2}c(t).
\eea
We are interested in the transmission of a single input photon.
The transmitted photon consists of elastic and inelastic components.
Thus, the state vector evolves on transmission as
\beq
a^{\dag}|1\ra \to t_e a^{\dag}|1\ra+t_i a^{\dag}e^{\dag}|1\ra,
\label{eq:tr1}
\eeq
where $e^{\dag}$ denotes an environmental excitation near the qubit.
As we show in Appendix~B1,
the fidelity and success probability of our EO are maximized 
when the spectral width of the input photon is 
much narrower than the cavity linewidth (i.e., the long pulse limit).
We thus assume the long pulse limit in the remainder of the paper.
The coefficients $t_e$ and $t_i$ are then determined
by considering the linear response to a classical continuous wave.
Setting $a_{in}=Ee^{-i\om_ct}$ and $a'_{in}=d_{in}=e_{in}=0$,
the dimensionless system variables
($x_c=i\sqrt{\kap/2}\ \la c \ra/a_{in}$,
$x_{\s}=-\sqrt{\kap/2}\ \la\s \ra/a_{in}$, and 
$x_{c^{\dag}c}=\kap\la c^{\dag}c \ra/2|a_{in}|^2$) are given by
\bea
x_c &=& 
\frac{\kap(\gam/2+\gam_p+i\Delta)}{\kap(\gam/2+\gam_p+i\Delta)+2g^2},
\label{eq:xc}
\\
x_{\s} &=& 
\frac{\kap g}{\kap(\gam/2+\gam_p+i\Delta)+2g^2},
\\
x_{c^{\dag}c} &=& 
\kap \frac{[\gam+2g^2{\rm Re}(1/\xi)]{\rm Re}(x_c)
-g\gam{\rm Re}(x_{\s}/\xi)}
{\kap\gam+2g^2(\kap+\gam){\rm Re}(1/\xi)},
\eea
where $\Delta=\om_q-\om_c$ is the detuning between the qubit and the input photon
and $\xi=(\kap+\gam)/2+\gam_p+i\Delta$.
$t_e$ and $t_i$ are related to the amplitude and flux transmissivities by
$t_e=\la a_{out}\ra/a_{in}$
and $|t_e|^2+|t_i|^2=\la a_{out}^{\dag}a_{out}\ra/|a_{in}|^2$.
Thus, we have
\bea
t_e &=& x_c,
\label{eq:te}
\\
|t_i|^2 &=& x_{c^{\dag}c}-|x_c|^2.
\eea
We can confirm that inelastic transmission
originates from pure dephasing, 
since $t_i=0$ when $\gam_p=0$.
The photon dynamics for the qubit state $|0\ra$
is obtained by taking the $g\to 0$ limit,
where we can confirm that $t_e=1$ and $t_i=0$.
Therefore, the counterpart of Eq.~(\ref{eq:tr1}) is 
\beq
a^{\dag}|0\ra \to a^{\dag}|0\ra.
\label{eq:tr0}
\eeq
Using these rigorous photon--qubit interactions
[Eqs.~(\ref{eq:tr1}) and (\ref{eq:tr0})],
we reconsider the time evolution of the initial state vector [Eq.~(\ref{eq:init})].
Since environmental excitation inhibits photon interference at BS2,
the state vector that clicks the detector is given by
\beq
|\psi\ra = 
a_L^{\dagger}\left(
\frac{t_e-1}{\sqrt{8}}|\phi_t\ra +
\frac{t_i}{\sqrt{8}}e_R^{\dagger}|\phi_2\ra -
\frac{t_i}{\sqrt{8}}e_L^{\dagger}|\phi_3\ra
\right), 
\eeq
where $|\phi_2\ra=(|01\ra+|11\ra)/\sqrt{2}$
and $|\phi_3\ra=(|10\ra+|11\ra)/\sqrt{2}$.
The click probability $\cP$, reduced density matrix $\hat{\rho}$,
and fidelity $\cF$ are respectively defined by 
$\cP=\la\psi|\psi\ra$, 
$\hat{\rho}={\rm Tr}_{a,b}\{|\psi\ra\la\psi|\}/\la\psi|\psi\ra$,
and $\cF =\la\phi_t|\hat{\rho}|\phi_t\ra$.
$\cF$ and $\cP$ are given by
\bea
\cF &=& \frac{|1-t_e|^2+|t_i|^2/2}{|1-t_e|^2+2|t_i|^2},
\label{eq:fidelity}
\\
\cP &=& |1-t_e|^2/8+|t_i|^2/4.
\label{eq:prob}
\eea

We first examine the effects of homogeneous broadening
by assuming that 
both detuning and spontaneous emission are absent
($\Delta=\gamma=0$).
In this case, the transmission probability through the cavity
is given by $|t_e|^2+|t_i|^2=\kap\gam_p/(\kap\gam_p+2g^2)$.
Therefore, when $\kappa\gamma_p/g^2 \ll 1$,
the cavity nearly completely suppresses transmission of the photon
and the present scheme functions with a high fidelity.
To achieve $\cF > 0.9$ (0.95),
$\kappa\gamma_p/g^2$ should be less than 0.15 (0.07).
Consequently, high-Q cavities satisfying $\kappa\gamma_p/g^2\ll 1$ are required to achieve high-fidelity EOs under a large homogeneous broadening.
This is qualitatively consistent with another scheme 
that employs resonant input photons~\cite{CTSL01a}.

Spontaneous emission usually degrades 
the figure of merits of quantum devices. 
In contrast,
spontaneous emission makes our protocol more robust 
against environmental noise and relaxes the cavity conditions,
so that a high-fidelity EO 
becomes possible between homogeneously broadened matter qubits
even in the weak-coupling regime ($g<\kap,\gam,\gam_p$), as
we show in Appendix~B2. 
The origin of infidelity here is inelastic scattering
(i.e., entanglement with the environment) that occurs 
while the matter qubit is being excited. 
Spontaneous emission reduces the lifetime of the excited state
and thus hinders inelastic scattering. 
However, in actual experiments, 
it is difficult to artificially increase the spontaneous emission rate
and thus this does not provide a practical solution.
Therefore, we look for another way to suppress environmental noise 
using existing technology. We consider the use of detuning.

\begin{figure}[h]
 \includegraphics[width=130mm]{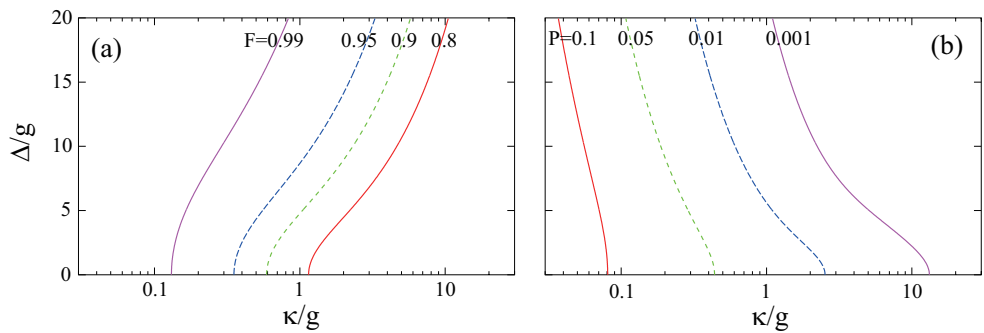}
 \caption{Contour plots of (a) fidelity and (b) success probability
 as functions of the cavity decay rate $\kappa$ and the detuning
  $\Delta$ in units of the cavity coupling strength $g$.
$\gam_p=\gam=2g$. 
Even when $\kappa=2g$, high-fidelity operation ($\cF=0.9$) 
is possible by setting $\Delta=9g$. 
The success probability will then be $\cP=0.13$\%.
}
 \label{fig:kD}
 \end{figure}

We briefly explain the physical mechanism of an EO
employing a detuned photon.
When there is large detuning $\Delta$,
Eqs.~(\ref{eq:xc}) and (\ref{eq:te}) give
$t_e \simeq e^{-i\theta}$, where $\theta=g^2/\Delta\kappa$.
Namely, when the qubit state is $|1\ra$,
the input photon acquires a phase shift 
that is determined by the product of the dispersive interaction ($g^2/\Delta$)
and the cavity photon lifetime ($\kappa^{-1}$)~\cite{matsuzaki2011entangling}.
This mechanism contrasts with that of resonant cases ($\Delta=0$),
where the transmitted wave is attenuated 
($t_e<1$) through scattering or reflection. 
The fidelity can be drastically improved by detuning $\Delta$
because detuning hinders real excitation of the matter qubit
and the resultant inelastic scattering.
Figure~\ref{fig:kD} shows a plot of the fidelity ($\cF$) and the success probability ($\cP$)
as functions of $\kappa$ and $\Delta$,
assuming $\gam=\gam_p=2g$.
The cavity condition for achieving $\cF=0.9$ is 
$\kap=0.59g$ when $\Delta=0$. 
However, this condition is relaxed to $\kap=2g$ by setting $\Delta=9g$.
Surprisingly, high-fidelity entanglement generation
is possible between homogeneously broadened matter qubits
even in the weak-coupling regime satisfying $g<\kap,\gam,\gam_p$.
Figure~\ref{fig:kD}(b) shows
that detuning reduces the success probability.
Namely, there is a trade-off between the fidelity and the success probability.
However, the success probability is $\cP=0.13$\%
when $\kap=2g$ and $\Delta=9g$,  
which is sufficiently large for practical use. 
The dark count rate is typically less than $10^{-7}$ per
nanosecond so that this success probability can exceed 
the dark count rate even within current technology. 

Finally, we describe a possible experimental realization
of our scheme using NV centers.
In a cavity QED setup composed of a diamond NV center and a microtoroidal cavity, 
the parameters $g$, $\kappa$, and $\gamma$
have comparable values of the order of tens of MHz \cite{park2006cavity},
while $\gamma_p$ is highly sensitive to temperature
\cite{fu2009observation}.
The linewidth of the NV center will be almost lifetime limited 
and thus $\gamma_p$ will be negligible at low temperatures such as $7$ K, 
whereas $\gamma_p$ will dominate the other parameters at higher temperatures.
At a low temperature ($\gamma_p=0.1g$ and $\gamma=g$),
$\cF=0.96$ can be attained with $\cP=0.13\%$
even by a low Q cavity
($\kap=4g$ and $\Delta =5g$).
Thus, using our scheme, it should be possible to realize an EO using current technology.
Moreover, even at higher temperatures, it should be possible
to perform an EO by our scheme
with modest requirements that are expected to be achievable in the near future.
Here, we set the parameters as $\gamma _p/2\pi = 300$ MHz (which
corresponds to a temperature of about $30$ K \cite{fu2009observation}), $\gamma /2\pi =
20$ MHz, $g/2\pi=250$ MHz, $\kappa /2\pi =150$ MHz, and $\Delta /2\pi =3$
GHz. Entanglement can be generated with $\cF=0.90$ and
$\cP=0.96\%$.
In principle,
once this amount of remote entanglement is achieved between distant nodes,
one can realize scalable distributed quantum computation by using
purification techniques inside the nodes \cite{GCR01a,
fujii2012topologicaltolerant}. Therefore,
distributed quantum computation may be possible at temperatures
of tens of kelvins, which can easily be generated without using liquid 
helium~\cite{daibo2011characteristics, felder2010optimization}.

In conclusion, 
we performed a non-perturbative analysis
of an EO using a detuned photon 
as a mediator between optically active matter qubits.
We demonstrated that this scheme is extremely robust 
against environmental noise so that
entanglement can be generated between homogeneously broadened matter qubits
even in the weak-coupling regime, where 
damping parameters overwhelm the coherent coupling between the cavity and the qubit.
This result is particularly relevant for
realizing distributed quantum computation by using NV centers
at high temperatures of the order of tens of kelvins.
Our scheme provides a practical way to overcome the main 
obstacle of using NV centers at high temperatures, 
namely large homogeneous broadening.

The authors thank H. Kosaka and W. J. Munro for helpful discussions.
This work was supported in part by the Funding Program for World-Leading 
Innovative R\&D on Science and Technology (FIRST), 
KAKENHI (22241025, 23104710, and 22244035),
SCOPE (111507004),
and NICT Commissioned Research.

\appendix
\section{cavity-QED analysis of single-photon dynamics}

\subsection{Hamiltonian and initial state vector}
\begin{figure}[h]
\includegraphics[width=65mm]{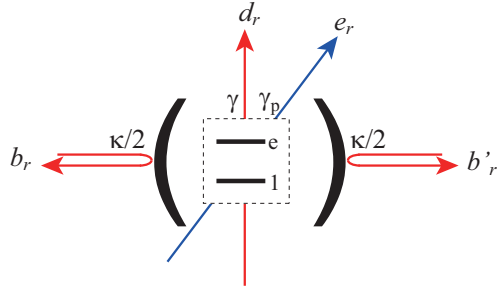}
\caption{The cavity-QED setup considered. 
A matter qubit is confined in a two-sided cavity,
and a single photon is input from the left-hand side.}
\label{fig:A1}
\end{figure}

We present here mathematical details on 
time evolution of a single input photon 
in the proposed optical circuit.
To begin with, we analyze transmission of a photon through a cavity.
The physical setup is illustrated in Fig.~\ref{fig:A1}.
It is composed of 
(i)~a matter qubit, which has three levels ($|0\ra$, $|1\ra$, $|e\ra$), 
(ii)~a two-sided cavity,
(iii)~leak fields from the cavity ($b$ and $b'$ fields),
(iv)~noncavity radiation modes ($d$ field), and 
(v)~environmental modes causing pure dephasing of the qubit ($e$ field).
Since the state $|0\ra$ is optically inactive,
we may regard the qubit as a two-level system ($|1\ra$, $|e\ra$)
when investigating its optical response.
Putting $\hbar=c=1$, the Hamiltonian is given by
\bea
\cH &=& \om_q\s^{\dg}\s+\om_c c^{\dg}c+g(\s^{\dg}c+c^{\dg}\s)
\nonumber
\\
&+& \int dk \left[ kb_k^{\dg}b_k + \sqrt{\kap/4\pi}(c^{\dg}b_k+b_k^{\dg}c) \right]
+ \int dk \left[ kb_k^{\prime\dg}b^{\prime}_k 
+ \sqrt{\kap/4\pi}(c^{\dg}b^{\prime}_k+b_k^{\prime\dg}c) \right]
\nonumber
\\
&+& \int dk \left[ kd_k^{\dg}d_k + \sqrt{\gam/2\pi}(\s^{\dg}d_k+d_k^{\dg}\s) \right]
+ \int dk \left[ ke_k^{\dg}e_k + \sqrt{\gam_p/\pi}\s^{\dg}\s(e_k+e_k^{\dg}) \right],
\label{eq:Ham}
\eea
where $\s(=|1\ra\la e|)$ and $c$ are  
the destruction operators of qubit and cavity photon,
and $\alp_k$ ($\alp=b$, $b'$, $d$, $e$) is the destruction operator
of $\alp$ field in the wavenumber representation.
The meanings of the parameters are given in the main text.
The field operator in the real-space representation
is defined by $\widetilde{\alp}_r=(2\pi)^{-1/2}\int dk e^{ikr}\alp_k$.
The $r<0$ ($r>0$) region corresponds to the incoming (outgoing) field.

At the initial moment ($t=0$),
we assume that a single photon is input from the $b$ field
and all other components are in their ground state.
The initial state vector is then written as
\beq
|\Psi_{in}\ra = \int dr f(r)\tb_r^{\dg}|1\ra,
\eeq
where $f(r)$ is the wavefunction of the input photon.
It is assumed to be
\beq
f(r)=\sqrt{2/l}\ \theta(-r)\exp(i\om_pr+r/l),
\eeq
where $\theta(r)$ is the Heavyside step function.
Namely, the input photon has a pulse length $l$
and a central frequency $\om_p$.

\subsection{Heisenberg equations}
From the Hamiltonian of Eq.~(\ref{eq:Ham}),
the raw Heisenberg equation for $b_k$ is given by
$db_k/dt = -ikb_k-i\sqrt{\kap/4\pi}\ c$.
This can be formally solved as
$b_k(t) = b_k(0)e^{-ikt}-i\sqrt{\kap/4\pi}
\int_0^t d\tau c(\tau)e^{-ik(t-\tau)}$.
As the Fourier transform of this equation,
$\tb_r(t)$ is given by
\beq
\tb_r(t) = \tb_{r-t}(0)-i\sqrt{\kap/2}\theta(r)\theta(t-r)c(t-r).
\eeq
Similarly, we have
\bea
\tb'_r(t) &=& \tb'_{r-t}(0)-i\sqrt{\kap/2}\theta(r)\theta(t-r)c(t-r),
\label{eq:io2}
\\
\td_r(t) &=& \td_{r-t}(0)-i\sqrt{\gam}\theta(r)\theta(t-r)\s(t-r),
\\
\te_r(t) &=& \te_{r-t}(0)-i\sqrt{2\gam_p}\theta(r)\theta(t-r)\s^{\dg}(t-r)\s(t-r).
\eea
These equations are known as the input-output relations.
The Heisenberg equations for $\s$ and $c$ are given by
\bea
\frac{d}{dt}\s &=& 
-i\tom_q \s -ig(1-2\s^{\dg}\s)c
-i\sqrt{\gam}(1-2\s^{\dg}\s)\td_{-t}(0)
-i\sqrt{2\gam_p}[\te_{-t}^{\dg}(0)\s+\s \te_{-t}(0)],
\label{eq:dsdt}
\\
\frac{d}{dt}c &=& -i\tom_c c-ig\s-i\sqrt{\kap/2}[\tb_{-t}(0)+\tb'_{-t}(0)],
\label{eq:dcdt}
\eea
where $\tom_q=\om_q-i(\gam/2+\gam_p)$ and $\tom_c=\om_c-i\kap/2$.

In the main text, the input and output fields 
($a_{in}$, $a'_{in}$, $a_{out}$, $a'_{out}$)
are defined as shown in Fig.~2.
They are related to the $b$ and $b'$ fields as
$a_{in}(t)=\tb_{-t}(0)$, $a'_{in}(t)=\tb'_{-t}(0)$,
$a_{out}(t)=\tb_{+0}(t)$ and $a'_{out}(t)=\tb'_{+0}(t)$.
After making these replacements,
Eqs.~(3)--(6) of the main text are derived.

\subsection{Correlation functions}
We discuss here the following one-time correlation functions,
$\alp_q(t)=\la 1|\s(t)\s^{\dg}|1\ra$, 
$\alp_c(t)=\la 1|c(t)\s^{\dg}|1\ra$,
$\beta_q(t)=\la 1|\s(t)|\Psi_{in}\ra$ and 
$\beta_c(t)=\la 1| c(t)|\Psi_{in}\ra$.
Their initial conditions are given by
$\alp_q(0)=1$ and $\alp_c(0)=\beta_q(0)=\beta_c(0)=0$. 
From Eqs.~(\ref{eq:dsdt}) and (\ref{eq:dcdt}),
their equations of motion are given by
\bea
\frac{d}{dt}
\left[\begin{array}{c}
\alp_q(t) \\ \alp_c(t)
\end{array}\right]
&=&
\left[\begin{array}{cc}
-i\tom_q & -ig \\ 
-ig & -i\tom_c
\end{array}\right]
\left[\begin{array}{c}
\alp_q(t) \\ \alp_c(t)
\end{array}\right],
\label{eq:alp}
\\
\frac{d}{dt}
\left[\begin{array}{c}
\beta_q(t) \\ \beta_c(t)
\end{array}\right]
&=&
\left[\begin{array}{cc}
-i\tom_q & -ig \\ 
-ig & -i\tom_c
\end{array}\right]
\left[\begin{array}{c}
\beta_q(t) \\ \beta_c(t)
\end{array}\right]
+
\left[\begin{array}{c}
0 \\ -i\sqrt{\frac{\kap}{2}}f(-t)
\end{array}\right].
\label{eq:beta}
\eea
We denote the Laplace transform of $\alp_q(t)$ by
$\cL_{\alp_q}(z)=\int_0^{\infty} dt e^{-zt}\alp_q(t)$.
Then, the Laplace transforms of 
the above equations are given by
\bea
\left[\begin{array}{c}
\cL_{\alp_q}(z) \\ \cL_{\alp_c}(z)
\end{array}\right]
&=&
\frac{1}{(z-\lam_1)(z-\lam_2)}
\left[\begin{array}{c}
z+i\tom_c \\ -ig
\end{array}\right],
\\
\left[\begin{array}{c}
\cL_{\beta_q}(z) \\ \cL_{\beta_c}(z)
\end{array}\right]
&=&
\frac{-i\sqrt{\kap/l}}{(z-\lam_1)(z-\lam_2)(z-\lam_3)}
\left[\begin{array}{c}
-ig \\ z+i\tom_q
\end{array}\right],
\eea
where $\lam_1$ and $\lam_2$ are the two roots of 
$(z+i\tom_q)(z+i\tom_c)+g^2=0$ and $\lam_3=-1/l-i\om_p$.
The one-time correlation functions are obtained
by analyzing the poles of the above Laplace transforms.

Next, we proceed to discuss the two-time functions such as 
$\beta^{(2)}_{q}(t_1,t_2)=\la 1|\s(t_1)\s^{\dg}(t_2)\s(t_2)|\Psi_{in}\ra$ and 
$\beta^{(2)}_{c}(t_1,t_2)=\la 1| c(t_1)\s^{\dg}(t_2)\s(t_2)|\Psi_{in}\ra$,
where $t_1>t_2$.
Their equations of motion with respect to $t_1$ 
are the same as Eq.~(\ref{eq:alp}),
and the initial conditions ($t_1\to t_2$) are given by
$\beta^{(2)}_{q}(t_2,t_2)=\beta_{q}(t_2)$ and 
$\beta^{(2)}_{c}(t_2,t_2)=0$.
Therefore, we have
\bea
\beta^{(2)}_{q}(t_1,t_2) &=& \alp_q(t_1-t_2)\beta_{q}(t_2),
\\
\beta^{(2)}_{c}(t_1,t_2) &=& \alp_c(t_1-t_2)\beta_{q}(t_2).
\eea
Repeating the same logic, 
general multi-time functions are written as
the products of one-time functions as
\bea
\beta^{(n)}_{q}(t_1,\cdots,t_n) &=& 
\alp_q(t_1-t_2)\alp_q(t_2-t_3)\cdots\beta_{q}(t_{n-1}-t_n),
\\
\beta^{(n)}_{c}(t_1,\cdots,t_n) &=& 
\alp_c(t_1-t_2)\alp_q(t_2-t_3)\cdots\beta_{q}(t_{n-1}-t_n).
\eea

\subsection{Wavefunctions of transmitted photon}
After interaction with the qubit-cavity system,
the input photon is reflected into the $b$ field,
transmitted into the $c$ field, or scattered into the $d$ field.
Time evolution of the input photon is determined by
$|\Psi(t)\ra=e^{-i\cH t}|\Psi_{in}\ra$.
The state vector of the transmitted component of photon is written as
\beq
|\Psi_c(t)\ra= \left[
\int dr g_0(r,t)\tb_r^{\prime\dg}
+\int drdx_1 g_1(r,x_1,t)\tb_r^{\prime\dg}\te_{x_1}^{\dg}
+\cdots \right]|1\ra.
\label{eq:psict1}
\eeq
Note that $0<r<x_1<\cdots<t$.
$g_0$ describes the elastic component,
whereas $g_n$ ($n\geq 1$) describes the inelastic component
that is entangled with the environmental modes ($e_x^{\dg}$).
We can determine $g_0$ as follows:
$g_0(r,t)=\la 1|\tb'_r|\Psi_c(t)\ra=
\la 1|\tb'_r(t)|\Psi_{in}\ra=-i\sqrt{\kap/2}\beta_c(t-r)$,
where Eq.~(\ref{eq:io2}) is used to derive the last equality.
Repeating the same arguments, we have
\bea
g_0(r,t) &=& -i\sqrt{\kap/2}\beta_c(t-r),
\\
g_1(r,x_1,t) &=& (-i\sqrt{\kap/2})(-i\sqrt{2\gam_p})\beta_q(t-x_1)\alp_c(x_1-r),
\\
g_n(r,x_1,\cdots,x_n,t) &=& (-i\sqrt{\kap/2})(-i\sqrt{2\gam_p})^n
\beta_q(t-x_n)\alp_q(x_n-x_{n-1})\cdots\alp_c(x_1-r).
\eea

On the other hand, when the qubit is in $|0\ra$,
the photon does not interact with the qubit
and inelastic processes are absent accordingly.
The state vector of the transmitted photon is then written as
\bea
|\overline{\Psi}_c(t)\ra &=& \int dr \overline{g}_0(r,t)\tb_r^{\prime\dg}|0\ra,
\label{eq:psict2}
\eea
where $\overline{g}_0(r,t)=\lim_{g\to 0}g_0(r,t)$.

\subsection{Fidelity and success probability}
Here we investigate the density matrix of 
matter qubits after an entanglement operation.
Throughout this section, we denote the photon field operator 
in the left (right) arm of the interferometer by $a_{Lr}$ ($a_{Rr}$).
The initial state vector is 
$|\psi_i\ra=2^{-1}\int dr f(r)a_{Lr}^{\dg}[|00\ra+|01\ra+|10\ra+|11\ra]$.
The beamsplitters divide a photon as
$a_{Lr}^{\dg} \to (ia_{Lr}^{\dg}+a_{Rr}^{\dg})/\sqrt{2}$ and 
$a_{Rr}^{\dg} \to (ia_{Rr}^{\dg}+a_{Lr}^{\dg})/\sqrt{2}$,
and the qubit-cavity system transforms a photon as
Eqs.~(\ref{eq:psict1})--(\ref{eq:psict2}).
When the photon is output in the left port of BS2,
the state vector of the overall system is given by
\bea
|\Psi_L\ra &=& \frac{1}{\sqrt{8}} \int dr 
\left[\overline{g}_0(r,t)-g_0(r,t)\right]\ta_{Lr}^{\dg}|\phi_t\ra,
\\
&-& \frac{1}{\sqrt{8}}
\sum_{n=1}^{\infty}\int drdx_1\cdots dx_n g_n(r,x_1,\cdots,x_n,t)
\ta_{Lr}^{\dg}\te_{Rx_1}^{\dg}\cdots\te_{Rx_n}^{\dg}|\phi_{e1}\ra,
\\
&+& \frac{1}{\sqrt{8}}
\sum_{n=1}^{\infty}\int drdx_1\cdots dx_n g_n(r,x_1,\cdots,x_n,t)
\ta_{Lr}^{\dg}\te_{Lx_1}^{\dg}\cdots\te_{Lx_n}^{\dg}|\phi_{e2}\ra,
\eea
where $|\phi_t\ra=(|01\ra-|10\ra)/\sqrt{2}$ is the target entangled state,
$|\phi_{e1}\ra=(|01\ra+|11\ra)/\sqrt{2}$,
and $|\phi_{e2}\ra=(|10\ra+|11\ra)/\sqrt{2}$.

The success probability $\cP$ of the entanglement operation,
namely, the probability to click the detector,
is given by $\cP=\la\psi_L|\psi_L\ra$. 
Denoting the norm of a function $f$ by $\cN(f)$, we have
\beq
\cP = \cN(\overline{g}_0-g_0)/8+\sum_{n=1}^{\infty}\cN(g_n)/4.
\eeq
The reduced density matrix $\rho$ of matter qubits is defined by
$\rho={\rm Tr}_{a,e}|\psi_L\ra\la\psi_L|/\la\psi_L|\psi_L\ra$.
Therefore,
\beq
\rho = \frac{\cN(\overline{g}_0-g_0)|\phi_t\ra\la\phi_t|
+\sum_{n=1}^{\infty}\cN(g_n)(|\phi_{e1}\ra\la\phi_{e1}|+|\phi_{e2}\ra\la\phi_{e2}|)}
{\cN(\overline{g}_0-g_0)+2\sum_{n=1}^{\infty}\cN(g_n)}.
\eeq
The fidelity $\cF$ between $\rho$ and the target state 
$|\phi_t\ra\la\phi_t|$ is given by
\beq
\cF = \frac{\cN(\overline{g}_0-g_0)+\sum_{n=1}^{\infty}\cN(g_n)/2}
{\cN(\overline{g}_0-g_0)+\sum_{n=1}^{\infty}2\cN(g_n)}.
\eeq
It is of note that the infinite sum of $\sum_{n=1}^{\infty}\cN(g_n)$
can be carried out analytically.
Using the Laplace transforms of $|\alp_c|^2$, $|\alp_q|^2$ and $|\beta_q|^2$,
we have
\beq
\sum_{n=1}^{\infty}\cN(g_n) = 
\kap\gam_p\frac{\cL_{|\beta_q|^2}(0)\cL_{|\alp_c|^2}(0)}{1-2\gam_p\cL_{|\alp_q|^2}(0)}.
\eeq
In the long pulse limit of $l\to\infty$,
$\cN(\overline{g}_0-g_0)$ and $\sum_{n=1}^{\infty}\cN(g_n)$
respectively reduce to $|1-t_e|^2$ and $|t_i|^2$
as discussed in the main text.

\section{numerical results}
In this section we present the numerical results 
that are not presented in the main text.
We assume $\om_p=\om_c$ throughout this section
and denote the qubit-cavity detuning $\om_q-\om_c$ by $\Delta$.
\subsection{Pulse Length}
\begin{figure}[h]
\begin{center}
\includegraphics[width=130mm]{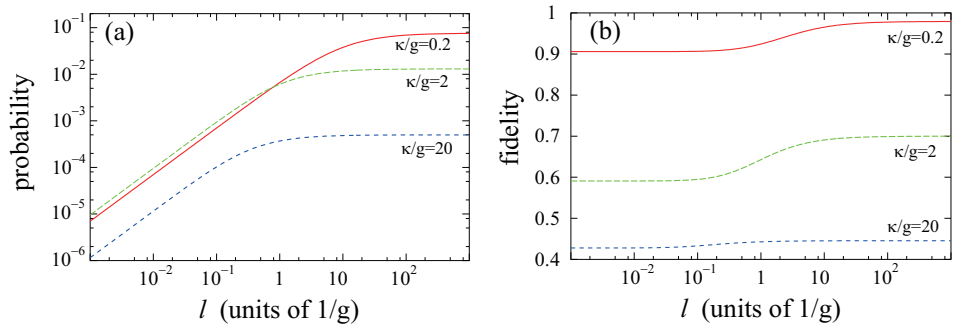}
\caption{Dependences of (a)~success probability and (b)~fidelity
on the input pulse length $l$.
$\gam=\gam_p=0$ and $\Delta=0$.
The values of $\kappa$ are indicated in the figure.}
\label{fig:ldep}
\end{center}
\end{figure}

First, we observe the effects of a finite pulse length $l$ of an input photon.
Assuming a dissipation-free ($\gam=\gam_p=0$) and resonant ($\Delta=0$) case,
the success probability $\cP$
is plotted as a function of $l$ for several values of $\kap$
in Fig.~\ref{fig:ldep}(a).
We can observe there that 
$\cP$ becomes independent of $l$ for $l\gg \kap^{-1}$
and reaches the limit value given by Eq.~(16) of main text.
This implies that the long-pulse limit,
where the input photon can enter the cavity perfectly,
is achieved when the spectral width of input photon ($l^{-1}$) 
is much narrower than that of cavity ($\kap$).
For shorter pulses,
the cavity filters out the off-resonant components of input photon
and the success probability is decreased accordingly.
In the short-pulse region,
the success probability becomes proportional to $l$
since it is determined 
by the overlap between the spectra of input photon and cavity.

Figure~\ref{fig:ldep}(b) shows the $l$-dependence of fidelity. 
As expected, $\cF$ becomes independent of $l$ in the long-pulse limit
and the limit value is given by Eq.~(15) of the main text.
However, in contrast with Fig.~\ref{fig:ldep}(a),
the fidelity is insensitive to $l$ also in the short pulse region.
This can be understood intuitively as follows.
Once the photon enters the cavity,
its property is determined by the cavity linewidth
and becomes irrelevant to the original linewidth determined by $l$.
We can observe that both the success probability and the fidelity
are maximized in the long pulse limit. 

\subsection{Spontaneous Emission}
\begin{figure}[h]
\includegraphics[width=130mm]{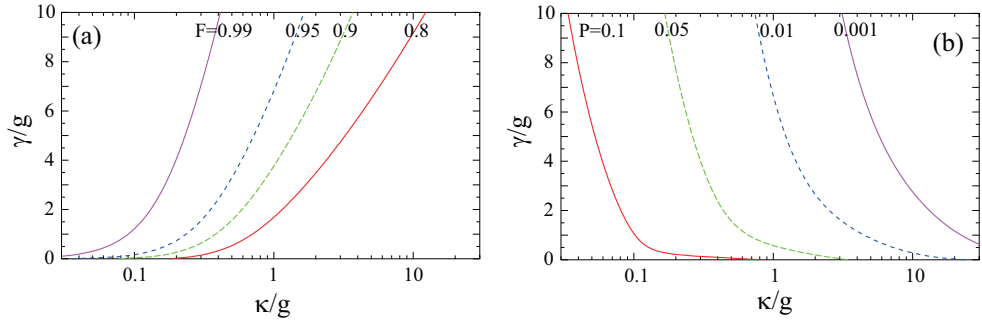}
\caption{Contour plots of (a) fidelity and (b) success probability,
as functions of $\kappa$ and $\gamma$.
$\gam_p=2g$ and $\Delta=0$.}
\label{fig:kr}
\end{figure}

Here we observe the effects of nonzero $\gamma$.
Assuming a noisy environment ($\gam_p=2g$) 
and a resonant input photon ($\Delta=0$),
the fidelity $\cF$ is plotted as a function of $\kappa$ and $\gamma$
in Fig.~\ref{fig:kr}(a).
We can confirm that the cavity condition
is substantially relaxed by a nonzero $\gamma$.
In order to achieve $\cF=0.9$ for example,
$\kap=0.08g$ is required when $\gamma$ is absent,
whereas this condition is relaxed to $\kap=0.59g$ when $\gam=2g$.
Usually, spontaneous emission into irrelevant modes 
leads to dissipation of quantum devices
and lowers their figure of merits.
However, this is not the case with the present scheme.
The origin of infidelity here is inelastic scattering
(in other words, entanglement with environment),
which occurs while the qubit is being excited.
Spontaneous emission makes the lifetime of excited state shorter
and thus hinders inelastic scattering.
The success probability $\cP$ is shown in Fig.~\ref{fig:kr}(b).
It is observed that $\cP$ is lowered by $\gamma$.
Thus, a high-fidelity operation becomes possible 
at the expense of a lower success probability.

\end{document}